\documentclass[aps,preprint]{revtex4}%
\usepackage{amsfonts}
\usepackage{amsmath}
\usepackage{amssymb}
\usepackage{graphicx}%
\begin{document}
\title{{\LARGE Shear and Vorticity in a Combined Einstein-Cartan-Brans-Dicke
Inflationary Lambda-Universe }\\Marcelo Samuel Berman$^{1}$}
\affiliation{$^{1}$Instituto Albert Einstein/Latinamerica\ - Av. Candido Hartmann, 575 -
\ \# 17}
\affiliation{80730-440 - Curitiba - PR - Brazil - email: msberman@institutoalberteinstein.org}
\date{24 December, 2007.}

\begin{abstract}
A combined BCDE (Brans-Dicke and Einstein-Cartan) theory \ with lambda-term is
developed through Raychaudhuri's equation, for inflationary scenario. It
involves a variable cosmological constant, which decreases with time, jointly
with energy density, cosmic pressure, shear, vorticity, and Hubble's
parameter, while the scale factor, total spin and scalar field increase
exponentially. The post-inflationary fluid resembles a perfect one, though
total spin grows, but the angular speed does not (Berman, 2007d).



\end{abstract}
\maketitle

\begin{center}

{\LARGE Shear and Vorticity in a Combined Einstein-Cartan-Brans-Dicke
Inflationary Lambda-Universe }\bigskip Marcelo Samuel Berman
\end{center}

\bigskip Berman(2007b), examined the time behavior of shear and vorticity in a
lambda-Universe, for inflationary models, in a Brans-Dicke framework. The
resulting scenario is that exponential inflation smooths the fluid, in order
to become a nearly perfect one after the inflationary period. We now examine
the inclusion of spin, by means of Einstein-Cartan's theory, when a scalar
field of Brans-Dicke origin, is included, \ along with a Cosmological lambda-term.

\bigskip{\large \bigskip}\bigskip Einstein-Cartan's gravitational theory,
though not bringing vacuum solutions different than those in General
Relativity theory, has an important r\^{o}le, by tying macrophysics, through
gravitational and electromagnetic phenomena (i.e., involving constants
\ $G$\ \ and $c$\ ), with microphysics, though Planck's constant, involving
spin originated by torsion. Intrinsic angular momentum was introduced by
Cartan as a Classical quantity (Cartan, 1923) before it was introduced as a
Quantum Theory element, around 1925. Of course, spin is important in the
Quantum Theory of particles. However, spin has taken part of Classical Field
Theory for a long time, and Cosmological models were treated as early as 1973
(Trautman, 1973). Einstein-Cartan Theory is the simplest Poincar\'{e} gauge
theory of gravity, in the frame of which, the gravitational field is described
by means of curvature and torsion, the sources being energy-momentum and spin
tensors. It is important to stress that torsion can be originated by spin but
not necessarily vice-versa.

\bigskip

\bigskip Though it was in the past, supposed that, due to spin,
Robertson-Walker's metric might not be representative of Physical reality in a
torsioned spacetime, recent papers recalled the approach shown by us in
several papers (Berman, 1990; 1991), on how anisotropic Bianchi-I models in
Einstein-Cartan's theory could be reduced to Robertson-Walker's prototype, by
defining overall, deceleration parameters, and scale-factors; we did the same
thing, with \ other papers dealing with anisotropic models in GRT and BD
theories [ for GRT see (Berman, 1988; Berman and Som, 1989 b); for BD theory
see (Berman and Som,1989) ].\ \ On the other hand, Berman and Som (2007) have
shown that, slight deviations from Robertson-Walker's metric, changing it to a
Bianchi-I metric, are enough to produce the anisotropic phenomena, like
entropy production, or other ones; this is a clue to the possibility of
considering overall scale-factors and deceleration parameters, etc, in the
Raychaudhuri's equation for Einstein-Cartan's Cosmology, without worrying with
any anisotropy, which becomes implicit in the equations of \ Raychaudhuri's
book (Raychaudhuri, 1979). The essential modification of General Relativistic
Bianchi-I cosmology, when we carry towards Einstein-Cartan's, resides, when
field equations are explicited, in that the normal energy momentum tensor
components \ \ $T_{1}^{1}$\ \ , \ $T_{2}^{2}$\ \ \ \ and\ \ $T_{3}^{3}$\ \ are
subtracted by a term \ \ $S^{2}$\ \ , while \ \ $T_{0}^{0}$\ \ is added by
\ \ $S^{2}$\ \ . Of course, there appear also non-diagonal \ $S-$\ dependent
terms: for instance, \ $T_{3}^{2}$\ \ and \ \ $T_{2}^{3}$\ \ depend linearly
with \ \ $S^{32}$\ . \ \ In our treatment of the Einstein-Cartan-Brans-Dicke
theory, the field equations are obviously satisfied, but we have short-cutted
the derivations, like we have done in the previous paper (Berman, 2007b),
which also conforms with the field equations of that case (Brans-Dicke theory
with lambda).\ \ The off-diagonal energy momentum components are null, for a
Robertson-Walker's framework.

\bigskip

\bigskip It is generally accepted that scalar tensor cosmologies play a
central r\^{o}le in the present view of the very early Universe (Berman,
2007). The cosmological "constant", which represents quintessence, is a time
varying entity, whose origin remounts to Quantum theory(Berman, 2007a). The
first, and most important scalar tensor theory was devised by Brans and
Dicke(1961), which is given in the "Jordan's frame". Afterwards, Dicke(1962)
presented a new version of the theory, in the "Einstein's frame", where the
field equations resembled Einstein's equations, but time, length, and inverse
mass, were scaled by a factor \ $\phi^{-\frac{1}{2}}$\ \ where \ $\phi
$\ \ stands for the scalar field. Then, the energy momentum tensor
\ \ $T_{ij}$\ \ is augmented\ by a new term $\Lambda_{ij}$\ , so that:

\bigskip

$G_{ij}=-8\pi G\left(  T_{ij}+\Lambda_{ij}\right)  $\ \ \ \ \ \ \ \ \ \ , \ \ \ \ \ \ \ \ \ \ \ \ \ \ \ \ \ \ \ \ \ \ \ \ \ \ \ \ \ \ \ \ \ \ \ \ \ \ \ \ \ \ \ \ \ \ \ \ \ \ \ (1)

\bigskip

where \ \ $G_{ij}$\ \ stands for Einstein's tensor. The new energy tensor
quantity, is given by:

\bigskip

$\Lambda_{ij}=\frac{2\omega+3}{16\pi G\phi^{2}}\left[  \phi_{i}\phi_{j}%
-\frac{1}{2}G_{ij}\phi_{k}\phi^{k}\right]  $ \ \ \ \ \ \ \ \ \ \ \ \ . \ \ \ \ \ \ \ \ \ \ \ \ \ \ \ \ \ \ \ \ \ \ \ \ \ \ \ \ \ \ \ \ \ \ \ \ \ (2)

\bigskip

In the above, \ $\omega$\ \ is the coupling constant. The other equation is:

\bigskip

$\square\log\phi=\frac{8\pi G}{2\omega+3}T$\ \ \ \ \ \ , \ \ \ \ \ \ \ \ \ \ \ \ \ \ \ \ \ \ \ \ \ \ \ \ \ \ \ \ \ \ \ \ \ \ \ \ \ \ \ \ \ \ \ \ \ \ \ \ \ \ \ \ \ \ \ \ \ \ \ \ \ \ \ \ \ (3)

\bigskip

where \ $\square$\ \ is the generalized d'Alembertian, and $T=T_{i}^{i}%
$\ \ \ .\ \ It is useful to remember that the energy tensor masses are also
scaled by \ $\phi^{-\frac{1}{2}}$\ \ .

\bigskip

For\ the Robertson-Walker's flat metric,

\bigskip

$ds^{2}=dt^{2}-\frac{R^{2}(t)}{\left[  1+\left(  \frac{kr^{2}}{4}\right)
\right]  ^{2}}d\sigma^{2}$ \ \ \ \ \ \ \ \ \ \ \ \ \ \ , \ \ \ \ \ \ \ \ \ \ \ \ \ \ \ \ \ \ \ \ \ \ \ \ \ \ \ \ \ \ \ \ \ \ \ \ \ \ \ \ \ \ \ (4)

\bigskip

where \ \ $k=0$\ \ and \ $d\sigma^{2}=dx^{2}+dy^{2}+dz^{2}$\ \ .

\bigskip

The field equations now read, in the alternative Brans-Dicke
reformulation(Raychaudhuri, 1979):

\bigskip

$\frac{8\pi G}{3}\left(  \rho+\frac{\Lambda}{\kappa}+\rho_{\lambda}\right)
=H^{2}\equiv\left(  \frac{\dot{R}}{R}\right)  ^{2}$ \ \ \ \ \ \ \ \ \ \ \ . \ \ \ \ \ \ \ \ \ \ \ \ \ \ \ \ \ \ \ \ \ \ \ \ \ \ \ \ \ \ \ \ \ \ \ \ \ (5)

\bigskip

$-8\pi G\left(  p-\frac{\Lambda}{\kappa}+\rho_{\lambda}\right)  =H^{2}%
+\frac{2\ddot{R}}{R}$ \ \ \ \ \ \ \ \ \ \ \ \ \ \ \ \ . \ \ \ \ \ \ \ \ \ \ \ \ \ \ \ \ \ \ \ \ \ \ \ \ \ \ \ \ \ \ \ \ \ (6)

\bigskip

In the above, we have: \ 

$\bigskip$

$\rho_{\lambda}=\frac{2\omega+3}{32\pi G}\left(  \frac{\dot{\phi}}{\phi
}\right)  ^{2}=\rho_{\lambda0}\left(  \frac{\dot{\phi}}{\phi}\right)  ^{2}$
\ \ \ \ \ \ \ . \ \ \ \ \ \ \ \ \ \ \ \ \ \ \ \ \ \ \ \ \ \ \ \ \ \ \ \ \ \ \ \ \ \ \ \ \ \ \ \ \ (7)

\bigskip

From the above equations (5), (6) and (7) we obtain:

\bigskip

$\frac{\ddot{R}}{R}=-\frac{4\pi G}{3}\left(  \rho+3p+4\rho_{\lambda}%
-\frac{\Lambda}{4\pi G}\right)  $ \ \ \ \ \ \ \ \ \ \ \ \ \ . \ \ \ \ \ \ \ \ \ \ \ \ \ \ \ \ \ \ \ \ \ \ \ \ \ \ \ \ \ \ \ (8)

\bigskip

Relation (8) represents Raychaudhuri's equation for a perfect fluid. By the
usual procedure, we would find the Raychaudhuri's equation in the general
case, involving shear ($\sigma_{ij}$) and vorticity ($\varpi_{ij}$); the
acceleration of the fluid is null for the present case, and then we find:

\bigskip

$3\dot{H}+3H^{2}=2\left(  \varpi^{2}-\sigma^{2}\right)  -4\pi G\left(
\rho+3p+4\rho_{\lambda}\right)  +\Lambda$ \ \ \ \ \ \ \ \ , \ \ \ \ \ \ \ \ \ \ \ (9)\ \ \ 

\bigskip

where \ $\Lambda$\ \ stands for a cosmological "constant". As we are mimicking
Einstein's field equations, $\Lambda$\ \ in (9) stands like it were a constant
(see however, Berman, 2007, 2007a, 2006b, 2006c). Notice that, \ when we
impose that the fluid is not accelerating, this means that the quadri-velocity
is tangent to the geodesics, i.e., the only interaction is gravitational.

\bigskip

\bigskip When Raychaudhuri's equation is calculated for non-accelerated fluid,
taken care of Einstein-Cartan's theory, combined with Brans-Dicke theory, the
following equation was found by us, based on the original calculation for
Einstein-Cartan's theory by Raychaudhuri (1979):

\bigskip

$3\dot{H}+3H^{2}=2\varpi^{2}-2\sigma^{2}-4\pi G\left(  \rho+3p+4\rho_{\lambda
}\right)  +\Lambda+128\pi^{2}S^{2}$ \ \ \ \ \ \ , \ \ (10)\ \ \ 

\bigskip

where \ \ $S$\ \ stands for the spin density contents of the fluid, where we
have omitted a term like

\bigskip

$\varpi S=\varpi_{ik}S^{ik}+\varpi^{ik}S_{ik}$%
\ \ \ \ \ \ \ \ \ \ \ \ \ \ \ \ \ \ \ \ \ , \ \ \ \ \ \ \ \ \ \ \ \ \ \ \ \ \ \ \ \ \ \ \ \ \ \ \ \ \ \ \ \ \ \ \ \ \ \ \ \ \ \ (10a)

\bigskip

which is to be included in the pressure and energy density terms, by a re-scaling.

\bigskip

It is important to stress, that relation (10) is the same general relativistic
equation, with the additional spin term, which transforms it into
Einstein-Cartan's equation. When we work a combined Einstein-Cartan's and
Brans-Dicke theory (BCDE theory), we would need to calculate the new field
equations for the combined theory.

\bigskip By employing the total action (Sabbata and Gasperini, 1985),

$\bigskip$

$L=\int d^{4}x\sqrt{-g}\left[  \pounds _{m}\left(  \psi,\triangledown
\psi,g\right)  -\frac{1}{2\chi}R\left(  g,\partial g,Q\right)  \right]  $
\ \ \ \ \ \ \ \ \ \ \ , \ \ \ \ \ \ \ \ (10b)

\bigskip

where the matter Lagrangian contains torsion because the connection is not
symmetric, and $\chi$\ \ is the coupling constant, both for curvature and
torsion, and when we perform independent variations with respect to \ $\psi
$\ \ , \ $g_{\mu\nu}$\ \ and \ $Q_{\mu\nu}^{\alpha}$\ \ \ ; the last the
one\ is the torsion tensor,

\bigskip

$Q_{\alpha\beta}^{\mu}=\frac{1}{2}\left(  \Gamma_{\alpha\beta}^{\mu}%
-\Gamma_{\beta\alpha}^{\mu}\right)  $ \ \ \ \ \ \ \ . \ \ \ \ \ \ \ \ \ \ \ \ \ \ \ \ \ \ \ \ \ \ \ \ \ \ \ \ \ \ \ \ \ \ \ \ \ \ \ \ \ \ \ \ \ \ \ \ \ \ \ \ (10c)

\bigskip

We find, the Einstein tensor,

\bigskip

$G^{\mu\nu}-\hat{\triangledown}_{\alpha}\left(  T^{\mu\nu\alpha}-T^{\nu
\alpha\mu}+T^{\alpha\mu\nu}\right)  =\chi T^{\mu\nu}$ \ \ \ \ \ \ , \ \ \ \ \ \ \ \ \ \ \ \ \ \ \ \ \ \ \ \ \ \ \ \ \ \ \ \ (10d)

\bigskip

where,

\bigskip

$T^{\mu\nu\alpha}=\chi S^{\mu\nu\alpha}$ \ \ \ \ \ \ \ \ \ \ \ .\ \ \ \ \ \ \ \ \ \ \ \ \ \ \ \ \ \ \ \ \ \ \ \ \ \ \ \ \ \ \ \ \ \ \ \ \ \ \ \ \ \ \ \ \ \ \ \ \ \ \ \ \ \ \ \ \ \ \ \ \ (10e)

\bigskip

We have defined,

\bigskip

$\hat{\triangledown}_{\alpha}\equiv\triangledown_{\alpha}+2Q_{\alpha
}=\triangledown_{\alpha}+2Q_{\alpha\nu}^{\nu}$ \ \ \ \ \ \ \ \ \ \ , \ \ \ \ \ \ \ \ \ \ \ \ \ \ \ \ \ \ \ \ \ \ \ \ \ \ \ \ \ \ \ \ \ \ \ \ (10f)

\bigskip

while the modified torsion tensor,

\bigskip

$T_{\mu\nu}^{\alpha}=Q_{\mu\nu}^{\alpha}+\delta_{\mu}^{\alpha}Q_{\nu}%
-\delta_{\nu}^{\alpha}Q_{\mu}$ \ \ \ \ \ \ \ \ \ \ \ \ \ \ . \ \ \ \ \ \ \ \ \ \ \ \ \ \ \ \ \ \ \ \ \ \ \ \ \ \ \ \ \ \ \ \ \ \ \ \ \ (10g)

\bigskip

\bigskip The resulting equations for a perfect fluid, can be found in
Raychaudhuri (1979):

\bigskip$-8\pi p=\left[  \text{Brans-Dicke alternative Riemann tensor \ }%
G_{i}^{i}\text{\ }\right]  +256\pi^{2}S^{2}$ \ \ , \ \ \ (10h)

\bigskip$8\pi\rho=\left[  \text{Brans-Dicke alternative Riemann tensor
\ }G_{0}^{0}\text{\ }\right]  +256\pi^{2}S^{2}$ \ \ \ \ \ , \ \ (10i)

It is important to acknowledge, that the above field equations should be
applied into the pseudo-General Relativistic equations, i.e., the Brans-Dicke
alternative (unconventional) framework. A plausibility reasoning that
substitutes an otherwise lengthy \ calculation, is the following: the term
with spin, as well as it is added to the other general relativistic terms in
equation (10), should be added equally to equation \ (9), because this is the
Brans-Dicke equation in a general relativistic format. This equation is
written in the unconventional format (Dicke, 1962), i.e., the alternative
system of equations. We could not write so simply equation\ (10)\ if the terms
in it were those of conventional Brans-Dicke theory.

\bigskip

Consider now exponential inflation, like we find in Einstein's theory:

\bigskip

$R=R_{0}e^{Ht}$\ \ \ \ \ \ \ \ , \ \ \ \ \ \ \ \ \ \ \ \ \ \ \ \ \ \ \ \ \ \ \ \ \ \ \ \ \ \ \ \ \ \ \ \ \ \ \ \ \ \ \ \ \ \ \ \ \ \ \ \ \ \ \ \ \ \ \ \ \ \ \ \ \ \ \ \ \ \ \ \ \ (11)

\bigskip

and, \ as usual in General Relativity inflationary models,

\bigskip

$\Lambda=3H^{2}$ \ \ \ \ \ \ \ \ \ \ . \ \ \ \ \ \ \ \ \ \ \ \ \ \ \ \ \ \ \ \ \ \ \ \ \ \ \ \ \ \ \ \ \ \ \ \ \ \ \ \ \ \ \ \ \ \ \ \ \ \ \ \ \ \ \ \ \ \ \ \ \ \ \ \ \ \ \ \ \ \ \ \ (12)

\bigskip

For the time being, $H$ \ is just a constant, defined by \ $H=\frac{\dot{R}%
}{R}$\ \ . We shall see, when we go back to conventional\ \ Brans-Dicke
theory, that \ $H$\ \ is not the Hubble's constant.

\bigskip

From (11), we find $H=H_{0}=$\ constant.\ 

\bigskip

A\ solution of Raychaudhuri's equation\ (10), would be the following:

\bigskip

\bigskip$\sigma=\sigma_{0}e^{-\frac{\beta}{2}t}$ \ \ \ ;

\bigskip

$\varpi=\varpi_{0}e^{-\frac{\beta}{2}t}$ \ \ \ ;

\bigskip

$\rho=\rho_{0}e^{-\beta t}$ \ \ \ ;

\bigskip

$p=p_{0}e^{-\beta t}$ \ \ \ ; \ \ \ \ \ \ \ \ \ \ \ \ \ \ \ \ \ \ \ \ \ \ \ \ \ \ \ \ \ \ \ \ \ \ \ \ \ \ \ \ \ \ \ \ \ \ \ \ \ \ \ \ \ \ \ \ \ \ \ \ \ \ \ \ \ \ \ \ \ \ \ \ \ \ \ (13)

\bigskip

$\phi=\phi_{0}e^{-\frac{\beta}{2}\sqrt{A}\text{ \ }e^{-\frac{\beta}{2}t}}$ \ \ \ \ \ \ \ \ \ .

\bigskip

$\Lambda=\Lambda_{0}=$ constant.

\bigskip

\bigskip$S_{U}=SR^{3}=s_{0}R_{0}^{3}e^{Ht}$ \ \ \ \ \ \ \ \ \ \ \ \ .

\bigskip In the above, \ $\sigma_{0}$\ , \ $\phi_{0}$\ ,\ \ $p_{0}$\ \ ,
\ \ $\rho_{0}$\ , $\beta$\ , \ $s_{0}$\ and \ $R_{0}$\ , are constants, and,
\ \ $S_{U}$\ \ stands for the total spin of the Universe, whose spin density equals,

\bigskip

$S=s_{0}e^{-\frac{\beta}{2}t}=s_{0}e^{-2Ht}$\ \ \ \ \ \ \ \ \ , \ \ \ \ \ \ \ \ \ \ \ \ \ \ \ \ \ \ \ \ \ \ \ \ \ \ \ \ \ \ \ \ \ \ \ \ \ \ \ \ \ \ \ \ \ \ \ \ \ \ \ \ (14)

\bigskip

while,

\bigskip

$\beta=4H$\ \ \ \ \ . \ \ \ \ \ \ \ \ \ \ \ \ \ \ \ \ \ \ \ \ \ \ \ \ \ \ \ \ \ \ \ \ \ \ \ \ \ \ \ \ \ \ \ \ \ \ \ \ \ \ \ \ \ \ \ \ \ \ \ \ \ \ \ \ \ \ \ \ \ \ \ \ \ \ \ \ \ (15)

\bigskip

\bigskip The ultimate justification for this solution is that one finds a good
solution in the conventional units theory, and that the Universe must expand.

\bigskip

When we return to conventional units, we retrieve the following corresponding solution:

\bigskip

$\bar{R}=R_{0}\phi^{\frac{1}{2}}e^{Ht}$ \ \ \ \ \ \ \ \ ; \ \ \ \ \ \ \ \ \ \ \ \ \ \ \ \ \ \ \ \ \ \ \ \ \ \ \ \ \ \ \ \ \ \ \ \ \ \ \ \ \ \ \ \ \ \ \ \ \ \ \ \ \ \ \ 

\bigskip

$\bar{\rho}=\rho_{0}\phi^{-2}e^{-\beta t}$\ \ \ \ \ \ \ \ \ ;

\bigskip

$\bar{p}=p_{0}\phi^{-2}e^{-\beta t}=\left[  \frac{p_{0}}{\rho_{0}}\right]
\bar{\rho}$\ \ \ \ \ \ \ \ \ ;

\bigskip\ \ \ \ \ \ \ \ \ \ \ \ \ \ \ \ \ \ \ \ \ \ \ \ \ \ \ \ \ \ \ \ \ \ \ \ \ \ \ \ \ \ \ \ \ \ \ \ \ \ \ \ \ \ \ \ \ \ \ \ \ \ \ \ \ \ \ \ \ \ \ \ \ \ \ \ \ \ \ \ \ \ \ \ \ \ \ \ \ \ \ \ \ \ \ \ \ \ \ (16)

$\bar{\sigma}=\sigma\phi^{-\frac{1}{2}}$ \ \ \ \ \ \ \ \ \ \ \ \ \ \ ;

\bigskip

$\bar{\varpi}=\varpi\phi^{-\frac{1}{2}}$ \ \ \ \ \ \ \ \ \ \ \ \ \ \ ;

\bigskip

$\bar{\Lambda}=\Lambda_{0}\phi^{-1}$\ \ \ \ \ \ \ \ \ \ \ \ \ ;

\bigskip

\bigskip$\bar{\phi}=\phi=\phi_{0}e^{-\frac{\beta}{2}\sqrt{A}\text{
\ }e^{-\frac{\beta}{2}t}}$ \ \ \ \ \ \ \ \ \ .

\bigskip

\bigskip We also have,

\bigskip$\bar{S}_{U}=S_{U}=s_{0}R_{0}^{3}e^{Ht}$
\ \ \ \ \ \ ,\ \ \ \ \ in\ \ \ \ \ $c=1$\ \ \ units\ \ \ \ . \ \ \ \ \ \ \ \ \ \ \ \ \ \ \ \ \ \ \ \ (17)

\bigskip

As we promised to the reader, $H$\ is not the Hubble's constant. Instead, we find:

\bigskip

\bigskip$\bar{\Lambda}=\Lambda_{0}$ $\phi_{0}^{-1}$ $e^{\frac{\beta}{2}%
\sqrt{A}\text{ }e^{-\frac{\beta}{2}t}}$ \ \ \ \ ; \ \ \ \ \ \ \ \ \ \ \ \ \ \ \ \ \ \ \ \ \ \ \ \ \ \ \ \ \ \ \ \ \ \ \ \ \ \ \ \ \ \ \ \ \ \ \ \ \ \ \ \ \ \ \ \ (18)

\bigskip

$\bar{\rho}=\rho_{0}$ $\phi_{0}^{-2}$ $e^{\beta\left[  \sqrt{A}\text{
\ }e^{-\frac{\beta}{2}\text{ }t}-\text{ }t\right]  }$ \ \ \ \ \ \ ;\ \ \ \ \ \ \ \ \ \ \ \ \ \ \ \ \ \ \ \ \ \ \ \ \ \ \ \ \ \ \ \ \ \ \ \ \ \ \ \ \ \ \ \ \ \ \ \ \ \ (19)

\bigskip

$\bar{p}=p_{0\text{ }}\phi_{0}^{-2}$ $e^{\beta\left[  \sqrt{A}\text{
\ }e^{-\frac{\beta}{2}\text{ }t}-\text{ }t\right]  }$ \ \ \ \ \ \ ;\ \ \ \ \ \ \ \ \ \ \ \ \ \ \ \ \ \ \ \ \ \ \ \ \ \ \ \ \ \ \ \ \ \ \ \ \ \ \ \ \ \ \ \ \ \ \ \ \ \ \ (20)

\bigskip

$\bar{R}=R_{0}$ $\phi_{0}^{-\frac{1}{2}}$ $e^{\left[  H\text{ }t\text{ }%
-\frac{1}{4}\beta\text{ }\sqrt{A}\text{ \ }e^{-\frac{\beta}{2}\text{ }%
t}\right]  }$ \ \ \ \ \ \ ;\ \ \ \ \ \ \ \ \ \ \ \ \ \ \ \ \ \ \ \ \ \ \ \ \ \ \ \ \ \ \ \ \ \ \ \ \ \ \ \ \ \ \ \ (21)

\bigskip

\bigskip$\bar{\sigma}=\sigma_{0}$ $\phi_{0}^{-\frac{1}{2}}$ $e^{-\frac{1}%
{2}\beta\left[  \text{ }t\text{ }-\frac{1}{2}\text{ }\sqrt{A}\text{
\ }e^{-\frac{\beta}{2}\text{ }t}\right]  }$ \ \ \ \ \ \ ,\ \ \ \ \ \ \ \ \ \ \ \ \ \ \ \ \ \ \ \ \ \ \ \ \ \ \ \ \ \ \ \ \ \ \ \ \ \ \ \ \ \ \ \ (22)

\bigskip

$\bar{\varpi}=\varpi_{0}$ $\phi_{0}^{-\frac{1}{2}}$ $e^{-\frac{1}{2}%
\beta\left[  \text{ }t\text{ }-\frac{1}{2}\text{ }\sqrt{A}\text{ \ }%
e^{-\frac{\beta}{2}\text{ }t}\right]  }$ \ \ \ \ \ \ ,\ \ \ \ \ \ \ \ \ \ \ \ \ \ \ \ \ \ \ \ \ \ \ \ \ \ \ \ \ \ \ \ \ \ \ \ \ \ \ \ \ \ (23)

\bigskip

and,

\bigskip

$\bar{H}=H$ $\phi_{0}^{-\frac{1}{2}}$ $e^{\frac{1}{4}\beta\sqrt{A}\text{
\ }e^{-\frac{\beta}{2}\text{ }t}}>0$
\ \ \ \ \ \ .\ \ \ \ \ \ \ \ \ \ \ \ \ \ \ \ \ \ \ \ \ \ \ \ \ \ \ \ \ \ \ \ \ \ \ \ \ \ \ \ \ \ \ \ \ \ \ \ (23
a)

\bigskip

\bigskip The fluid obeys a perfect gas equation of state. It represents a
radiation phase, if we impose,

\bigskip

$p_{0}=\frac{1}{3}\rho_{0}$ \ \ \ \ \ \ \ \ \ \ \ \ \ . \ \ \ \ \ \ \ \ \ \ \ \ \ \ \ \ \ \ \ \ \ \ \ \ \ \ \ \ \ \ \ \ \ \ \ \ \ \ \ \ \ \ \ \ \ \ \ \ \ \ \ \ \ \ \ \ \ \ \ \ \ \ \ \ \ \ \ \ (24)

\bigskip

\bigskip Returning to Raychaudhuri's equation, we have the following condition
to be obeyed by the constants:

\bigskip

$\sigma_{0}^{2}-\varpi_{0}^{2}=-2\pi G\left[  \rho_{0}+3p_{0}+4\rho_{\lambda
0}\right]  +64\pi^{2}s_{0}^{2}$ \ \ \ \ \ \ \ . \ \ \ \ \ \ \ \ \ \ \ \ \ \ \ \ \ \ \ \ \ \ \ \ \ (25)

\bigskip

\bigskip We now investigate the limit when \ \ $t\longrightarrow\infty$\ \ of
the above formulae, having in mind that, by checking that limit, \ we will
know which ones increase or decrease with time; of course, we can not stand
with an inflationary period unless it takes only an extremely small period of
time. Remember that \ \ \ $\beta=4H>0$\ \ \ .

\bigskip

We find:

\bigskip

$\lim\limits_{t\longrightarrow\infty}\bar{H}=H\phi_{0}^{-1/2}$ \ \ \ \ ;\ \ \ 

\bigskip

$\lim\limits_{t\longrightarrow\infty}\bar{R}=\infty$ \ \ \ \ ;

\bigskip

$\lim\limits_{t\longrightarrow\infty}\bar{\sigma}=\lim
\limits_{t\longrightarrow\infty}\bar{\varpi}=0$ \ \ \ \ ;

\bigskip

$\lim\limits_{t\longrightarrow\infty}\bar{\rho}=\lim\limits_{t\longrightarrow
\infty}\bar{p}=0$ \ \ \ \ \ \ ;

\bigskip

$\lim\limits_{t\longrightarrow\infty}\bar{\Lambda}=\Lambda_{0}\phi_{0}^{-1}$ \ \ \ \ \ \ \ \ \ \ \ ;

\bigskip

$\lim\limits_{t\longrightarrow\infty}\bar{\phi}=\phi_{0}$ \ \ \ \ \ \ \ \ \ \ ;

\bigskip

\bigskip$\lim\limits_{t\longrightarrow\infty}\bar{S}_{U}=\infty$ \ \ \ \ .

\bigskip

By comparing the above limits, \ with the limit \ $t\rightarrow0$\ \ , as we
can check, the scale factor, total spin, and the scalar field, are
time-increasing, while all other elements of the model, namely, vorticity,
shear, Hubble's parameter, energy density, cosmic pressure, and cosmological
term, as described by the above relations, decay with time. This being the
case, shear and vorticity are decaying, so that, after inflation, we
retrieve\ \ a nearly perfect fluid: \ inflation has the peculiarity of
removing shear, and vorticity, but not spin, from the model. It has to be
remarked, that pressure and energy density obey a perfect gas equation of
state. The graceful exit from the inflationary period towards the early
Universe radiation phase, is attained with condition (24). We have found a
solution that is entirely compatible with the Brans-Dicke counterpart (Berman,
2007c). The total spin of the Universe grows, but the angular velocity does
not (Berman, 2007d).

\bigskip\bigskip

{\Large Acknowledgements}

\bigskip An anonymous referee made substantial contributions in order to
correct several inconveniencies in our submission, and the author recognizes
that those corrections were fundamental in order to bring a satisfactory
manuscript into publication. Many thanks to him.

\bigskip

The author also thanks his intellectual mentors, Fernando de Mello Gomide and
M. M. Som, and also to Marcelo Fermann Guimar\~{a}es, Nelson Suga, Mauro
Tonasse, Antonio F. da F. Teixeira, and for the encouragement by Albert, Paula
and Geni.

\bigskip

\bigskip{\Large References}

\bigskip Berman,M.S. (1988) - GRG \textbf{20}, 841.

Berman, M.S. (\bigskip1990) - Nuovo Cimento, \textbf{105B}, 1373.

Berman, M.S. (1991) - GRG, \textbf{23}, 1083.

\bigskip\bigskip Berman,M.S. (2006b) - \textit{Energy of Black-Holes and
Hawking's Universe \ }in \textit{Trends in Black-Hole Research, }Chapter
5\textit{.} Edited by Paul Kreitler, Nova Science, New York.

Berman,M.S. (2006c) - \textit{Energy, Brief History of Black-Holes, and
Hawking's Universe }in \textit{New Developments in Black-Hole Research},
Chapter 5\textit{.} Edited by Paul Kreitler, Nova Science, New York.

Berman,M.S. (2007) - \textit{Introduction to General Relativistic and Scalar
Tensor Cosmologies}, Nova Science, New York.

Berman,M.S. (2007a) - \textit{Introduction to General Relativity and the
Cosmological Constant Problem}, Nova Science, New York.

Berman,M.S. (2007c) - \textit{Shear and Vorticity in Inflationary Brans-Dicke
Cosmology with Lambda-Term, }Astrophysics and Space Science, \textbf{310, }205.

Berman,M.S. (2007d) - \textit{The Pioneer Anomaly and a Machian Universe,
}Astrophysics and Space Science, \textbf{312}, 275.

Berman,M.S.; Som, M.M. (1989) - Nuovo Cimento, \textbf{103B},N.2, 203.

Berman,M.S.;Som, M.M. (1989 b) - GRG , \textbf{21}, 967-970.

Berman,M.S.;Som, M.M. (2007) - \textit{Natural Entropy Production in an
Inflationary Model for a Polarized Vacuum}, Astrophysics and Space Science,
\textbf{310, }277. Los Alamos Archives:
http://www.arxiv.org/abs/physics/0701070 .

\bigskip Brans, C.; Dicke, R.H. (1961) - Physical Review, \textbf{124}, 925.

Cartan, E. (1923) - Ann. Ec. Norm. Sup., \textbf{40,} 325.

Dicke, R.H. (1962) - Physical Review, \textbf{125}, 2163.

\bigskip Raychaudhuri, A. K. (1979) - \textit{Theoretical Cosmology, }Oxford
University Press, Oxford.

Sabatta, V. de;\ Gasperini, M. (1985) - \textit{Introduction to Gravitation,
}World Scientific, Singapore.

Trautman, A. (1973) - Nature (Physical Science), \textbf{242,} 7.

\end{document}